\documentclass[conference]{IEEEtran}
\IEEEoverridecommandlockouts
\usepackage{cite}
\usepackage{amsmath,amssymb,amsfonts}
\usepackage{algorithmic,algorithm}
\usepackage{graphicx}
\usepackage{textcomp}
\usepackage{xcolor}
\usepackage{multirow} 
\def\BibTeX{{\rm B\kern-.05em{\sc i\kern-.025em b}\kern-.08em T\kern-.1667em\lower.7ex\hbox{E}\kern-.125emX}}

\ifCLASSOPTIONcompsoc
\usepackage[caption=false,font=normalsize,labelfont=sf,textfont=sf]{subfig}
\else
\usepackage[caption=false,font=footnotesize]{subfig}
\fi

\usepackage{cuted}

\newtheorem{theorem}{Theorem}

\newtheorem{lemma}{Lemma}

\newtheorem{corollary}{Corollary}

\newtheorem{definition}{Definition}

\renewcommand{\IEEEQED}{\IEEEQEDopen}

\usepackage{setspace}
\begin{document}
\bstctlcite{ref:BSTcontrol}

\makeatletter
\newcommand{\rmnum}[1]{\romannumeral #1}
\newcommand{\Rmnum}[1]{\expandafter\@slowromancap\romannumeral #1@}
\makeatother

\title{Intelligent Reflecting Surface Aided Multi-Cell NOMA Networks
	\thanks{This work was supported by the Beijing Natural Science Foundation (L182036), and the China Scholarship Council.}
}

\author{\IEEEauthorblockN{
		Wanli~Ni$^{*}$, 
		Xiao~Liu$^{\dag}$, 
		Yuanwei~Liu$^{\dag}$, 
		Hui~Tian$^{*}$, 
		Yue~Chen$^{\dag}$}
	\IEEEauthorblockA{
		$^{*}$State Key Lab. of Networking and Switching Technology, Beijing Univ. of Posts and Telecommun., Beijing, China \\ 
		$^{\dag}$School of Electronic Engineering and Computer Science, Queen Mary University of London, London, UK \\
		Email: \{charleswall,tianhui\}@bupt.edu.cn; \{x.liu,yuanwei.liu,yue.chen\}@qmul.ac.uk
	}
}

\maketitle

\begin{abstract}
	This paper proposes a novel framework of resource allocation in intelligent reflecting surface (IRS) aided multi-cell non-orthogonal multiple access (NOMA) networks, where a sum-rate maximization problem is formulated.
	To address this challenging mixed-integer non-linear problem, we decompose it into
	an optimization problem (P1) with continuous variables and a matching problem (P2) with integer variables.
	For the non-convex optimization problem (P1), iterative algorithms are proposed for allocating transmit power, designing reflection matrix, and determining decoding order by invoking relaxation methods such as convex upper bound substitution, successive convex approximation and semidefinite relaxation.
	For the combinational problem (P2), swap matching-based algorithms are proposed to achieve a two-sided exchange-stable state among users, BSs and subchannels.
	Numerical results are provided for demonstrating that the sum-rate of the NOMA networks is capable of being enhanced with the aid of the IRS.
\end{abstract}


\section{Introduction}
By modifying the amplitude and phase of reflective signals, the software-controlled intelligent reflecting surface (IRS) can reconfigure the wireless channels between the transmitters and receivers \cite{Liu2020Reconfigurable}.
This remarkable feature of IRS can be utilized to enhance the performance of wireless communication networks from various aspects such as coverage extension, secrecy improvement, and fairness guarantee \cite{Liu2020RIS}.
Compared to the conventional active relays supporting massive multiple-input multiple-output or millimeter-wave communication, decode or amplify are not requested in the IRS-aided wireless networks due to the reason that IRS is equipped with a large number of passive reflecting elements \cite{Wu2019IRS}.
Thus, both hardware cost and energy consumption of the IRS-aided wireless networks are lower than the conventional amplify-and-forward (AF) or decode-and-forward schemes.


Recently,
non-orthogonal multiple access (NOMA) has been deemed as a promising technique for enhancing network performance in terms of throughput and connectivity \cite{Liu2020NOMA},
where the successive interference cancellation (SIC) approach is adopted to decode informations at the receivers \cite{Hua2019Energy, Liu2017NOMA}.
More particularly, for the multi-cell NOMA networks with large-scale devices, the co-channel interference makes the resource allocation problem among base stations (BSs) coupled and correlated \cite{Cui2018QoE}, which leads to a challenging optimization problem.
Given these challenges, it is particularly important to jointly design user scheduling and resource allocation for performance improvement.

Inspired by the advantages of both IRS and NOMA, it is valuable and imperative to integrate them together to further improve the throughput, coverage and connectivity, due to the following reasons.
Firstly,
the interference can be suppressed by applying IRS into multi-cell NOMA networks and properly designing the reflection matrix of IRS \cite{Xie2020Max}.
Secondly,
for cell-edge NOMA users that suffers high signal attenuation, IRS can be deployed to passively relay the intended signal in a low-cost way, which is beneficial to provide better service for these cell-edge users with poor signal strength \cite{Wu2019IRS}.
Thirdly, 
the decoding order of NOMA users can be effectively tuned by adjusting the reflecting elements to reconfigure the propagation environment \cite{Ding2020IRS}. Therefore, IRS is also profitable to optimize the user pairing and connectivity.

At present, the majority of contributions on IRS-aided networks focus on theoretical analysis \cite{Yue2020Analysis} and performance optimization such as fairness \cite{Xie2020Max}, throughput \cite{Zuo2020Resource}, and efficiency \cite{Liu2020RIS, Huang2019EE}.
Considering the perfect and imperfect SIC of an IRS-aided NOMA network, Yue \textit{et. al} \cite{Yue2020Analysis} derived the exact expressions of outage probability and ergodic rate.
By alternatively optimizing the transmit and reflective beamforming, Xie \textit{et. al} \cite{Xie2020Max} maximized the minimal SINR at receivers to ensure user fairness.
According to the experimental results in \cite{Zuo2020Resource}, the throughput of NOMA networks can be further improved with the aid of IRS.
Compared to regular multi-antenna AF relays, Huang \textit{et. al} \cite{Huang2019EE} demonstrated that IRS-aided networks can enjoy a higher energy efficiency.
Although a few previous literatures on IRS-assisted NOMA networks have addressed the challenging transmit power and reflective beamforming problem iteratively, the system model is limited to single-cell and/or single-carrier setups.
To the best of our knowledge, this is the first work which addressing the resource allocation for IRS-aided multi-cell NOMA networks with multiple subchannels.
The contributions of this work are summarized as follows:
1) A mixed-integer non-linear problem in IRS-aided multi-cell NOMA networks is formulated to maximize the sum rate;
2) To solve the non-convex optimization subproblem, iterative algorithms are proposed for allocating transmit power, designing reflection matrix, and determining decoding order.
3) To tackle the combinational subproblem, swap matching-based algorithms are developed for achieving a two-sided exchange-stable state among users, BSs and subchannels.


\section{System Model and Problem Formulation}
\subsection{System Model}
We consider an IRS-aided multi-cell NOMA transmission scenario, where an IRS is deployed for enhancing wireless service from $J$ single-antenna BS to $I$ single-antenna cellular users, while $\mathcal{I}=\{1,2,\dots, I\}$ and $\mathcal{J}=\{1,2,\dots, J\}$.
The IRS is equipped with $M$ passive reflecting elements, denoted by $\mathcal{M}=\{1,2,\dots, M\}$.
The diagonal reflection matrix of IRS is denoted by $\mathbf{\Theta} = \text{diag} \left\lbrace e^{j\theta_1}, e^{j\theta_2}, \dots, e^{j\theta_M} \right\rbrace$, where $\theta_m \in [0,2\pi]$ denotes the phase shift of the $m$-th element on IRS.
The total bandwidth $W$ is divided into $K$ subchannels, denoted by $\mathcal{K}=\{1,2,\dots, K\}$, and all subchannels can be reused among BSs to improve the spectrum efficiency.
In an effort to reduce the decoding complexity of SIC procedure at the receivers, we assume that the number of paired NOMA users, simultaneously sharing the available spectrum in each cell, is no more than $A_{\text{max}}$, while $A_{\text{max}} \ge 2$.


Let $\alpha_{ij} \in \{0,1\}$ and $\beta_{jk} \in \{0,1\}$ denote the user association indicator and subchannel assignment factor, respectively.
Specifically, we have $\alpha_{ij}=1$ if the $i$-th user is associated with the $j$-th BS, otherwise $\alpha_{ij}=0$.
Furthermore, we have $\beta_{jk}=1$ if the $k$-th subchannel is assigned to the $j$-th BS, otherwise $\beta_{jk}=0$.
Hence, the $i$-th user will be served by the $j$-th BS on the $k$-th subchannel if and only if $\alpha_{ij}\beta_{jk}=1$, otherwise $\alpha_{ij}\beta_{jk}=0$.
Then, the superimposed signal, $x_{jk}$, broadcasted by the $j$-th BS on the $k$-th subchannel can be given by
\begin{equation}\label{superimposed_message}
x_{jk} = {\alpha_{ij}\beta_{jk} \sqrt{p_{ijk}} x_{ijk}} \ + \ { \sum \nolimits_{t \ne i} \alpha_{tj}\beta_{jk} \sqrt{p_{tjk}} x_{tjk}},
\end{equation}
where $x_{ijk}$ and $p_{ijk}$ denote the intended signal and power transmitted by BS $j$ on subchannel $k$ for user $i$, respectively.

Considering the intra-cell and inter-cell interference on the $k$-th subchannel, the received signal of user $i$ associated with BS $j$ on subchannel $k$ is expressed as
\begin{eqnarray}\label{received_signal}
y_{ijk}
&=&{\left( h_{ijk} + \mathbf{g}_{ik}^{\rm H} \mathbf{\Theta} \mathbf{f}_{jk}\right)  \alpha_{ij}\beta_{jk} \sqrt{p_{ijk}} x_{ijk}} \ + \  {z_{ijk}} \nonumber \\
&+&{\left( h_{ijk} + \mathbf{g}_{ik}^{\rm H} \mathbf{\Theta} \mathbf{f}_{jk}\right)  \sum \nolimits_{t \ne i} \alpha_{tj}\beta_{jk} \sqrt{p_{tjk}} x_{tjk}} \\
&+&{\sum \nolimits_{s \ne j} \left( h_{isk} + \mathbf{g}_{ik}^{\rm H} \mathbf{\Theta} \mathbf{f}_{sk}\right) \sum \nolimits_{i} \alpha_{is}\beta_{sk} \sqrt{p_{isk}} x_{isk}}, \nonumber
\end{eqnarray}
where
$h_{ijk}$ denotes the Rayleigh fading channel between BS $j$ and user $i$ on subchannel $k$,
$\mathbf{f}_{jk} \in \mathbb{C}^{M \times 1}$ represents the Rician fading channel between BS $j$ and IRS on subchannel $k$,
$\mathbf{g}_{ik} \in \mathbb{C}^{M \times 1}$ formulates the Rayleigh fading channel between IRS and user $i$ on subchannel $k$,
and $z_{ijk}$ is the additive white Gaussian noise (AWGN) with zero mean and variance $\sigma^2$.

When each user independently receives the superposition signal and decodes its desired message via SIC technique, we denote the SIC decoding order as $\pi_{jk}(i)$ for user $i$ associated with BS $j$ on subchannel $k$.
Specifically, we have $\pi_{jk}(i)=n$ if the message of user $i$ is the $n$-th signal to be decoded at the receiver, namely, user $i$ first decodes the signals of all the previous $(n-1)$ users, and then successively subtracts their signals to decode its own desired signal. 
For example, two users $i$ and $\tilde{i}$ associated with BS $j$ on subchannel $k$, satisfying $\pi_{jk}(i) \le \pi_{jk}(\tilde{i})$,
user $\tilde{i}$ is capable of successfully canceling interference from the superposition signal of user $i$ with the aid of SIC.
Let $H_{ijk}=h_{ijk} + \mathbf{g}_{ik}^{\rm H} \mathbf{\Theta} \mathbf{f}_{jk}$ denote the combined channel gain,
and $P_{ijk}=\alpha_{ij}\beta_{jk} p_{ijk}$ represents the power allocation.
Then, the decoding order constraints for guaranteeing success SIC can be formulated as equation (\ref{SIC_condition_1}) at the top of the next page.
It indicates that the achievable SINR of user $\tilde{i}$ to decode user $i$ is no less than that of user $i$.
By simple operations, (\ref{SIC_condition_1}) can be reformulated as (\ref{SIC_condition_2}) at the top of the next page.

\begin{figure*}[ht]
	\begin{equation}\label{SIC_condition_1}
	\frac{ | H_{\tilde{i}jk} | ^2 P_{ijk} }{ | H_{\tilde{i}jk} | ^2 \sum \limits_{ \pi_{jk}(\hat{i}) > \pi_{jk}(i) } P_{\hat{i}jk} + \sum \limits_{s=1, s \ne j}^{J} \left| H_{\tilde{i}sk} \right| ^2 \sum \limits_{i=1}^{I} P_{isk} + \sigma ^2 }
	\ge
	\frac{ | H_{ijk} | ^2 P_{ijk} }{ | H_{ijk} | ^2 \sum \limits_{ \pi_{jk}(\hat{i}) > \pi_{jk}(i) } P_{\hat{i}jk} + \sum \limits_{s=1, s \ne j}^{J} | H_{isk} | ^2 \sum \limits_{i=1}^{I} P_{isk} + \sigma ^2 },
	\end{equation}
	\begin{equation}\label{SIC_condition_2}
	\Delta_{jk}(i,\tilde{i}) =
	| H_{\tilde{i}jk} | ^2 \left( \sum \limits_{s=1, s \ne j}^{J} | H_{isk} | ^2 \sum \limits_{i=1}^{I} P_{isk} + \sigma ^2 \right)
	-
	| H_{ijk} | ^2 \left( \sum \limits_{s=1, s \ne j}^{J} | H_{\tilde{i}sk} | ^2 \sum \limits_{i=1}^{I} P_{isk} + \sigma ^2 \right) \ge 0, \text{if} \ \pi_{jk}(i) \le \pi_{jk}(\tilde{i}).
	\end{equation}
	\hrulefill
\end{figure*}

Accordingly, the received SINR of user $i$ associated with BS $j$ on subchannel $k$ is given by
\begin{eqnarray}\label{SINR}
\text{SINR}_{ijk} =
\frac{ \left| H_{ijk} \right| ^2 P_{ijk} }{ I_{ijk}^{\text{intra}} + I_{ijk}^{\text{inter}} + \sigma ^2 },
\end{eqnarray}
where
$I_{ijk}^{\text{intra}} = \left| H_{ijk} \right| ^2 \sum_{ \pi_{jk}(\hat{i}) > \pi_{jk}(i) } P_{\hat{i}jk}$
and
$I_{ijk}^{\text{inter}} = \sum_{s \ne j}\left| H_{isk} \right| ^2 \sum_{i} P_{isk}$
are the intra-cell and inter-cell interference, respectively.
Therefore, the corresponding achievable downlink data rate of user $i$ associated with BS $j$ on subchannel $k$ is calculated as
\begin{equation}\label{data_rate}
R_{ijk}=\frac{W}{K} \log_{2} \left( 1 + \frac{ \left| H_{ijk} \right| ^2 P_{ijk} }{ I_{ijk}^{\text{intra}} + I_{ijk}^{\text{inter}} + \sigma ^2 } \right).
\end{equation}

\subsection{Problem Formulation}
The objective of this paper is to maximize the sum rate, and the optimization problem can be formulated as
\begin{subequations}
\begin{eqnarray}
\label{variable}
&var  & \left\lbrace \alpha_{ij}, \beta_{jk}, \mathbf{\Theta}, p_{ijk}, \pi_{jk}(i) \ | \ \forall i,j,k \right\rbrace, \\
&\max & \sum \nolimits_{i} \sum \nolimits_{j} \sum \nolimits_{k} R_{ijk}, \\
&s.t. & \Delta_{jk}(i,\tilde{i}) \ge 0, \ \text{if} \ \pi_{jk}(i) \le \pi_{jk}(\tilde{i}), \\
&{}& \sum \nolimits_{j} \sum \nolimits_{k} R_{ijk} \ge R_{\text{min}}, \ \forall i, \\
&{}& \sum \nolimits_{i} \sum \nolimits_{k} P_{ijk} \le P_{\text{max}}, \ \forall j, \\
&{}& \sum \nolimits_{j} \alpha_{ij} = 1, \ \forall i, \\
&{}& 2 \le \sum \nolimits_{i} \alpha_{ij} \le A_{\text{max}}, \ \forall j, \\
&{}& \sum \nolimits_{k} \beta_{jk} \ge 1, \ \forall j, \\
&{}& \sum \nolimits_{j} \beta_{jk} \ge 1, \ \forall k, 
\end{eqnarray}
\end{subequations}
where the optimization variables are given in (\ref{variable}),
$R_{\text{min}}$ is the minimum data rate required by each user,
$P_{\text{max}}$ is the maximum transmission power provided by each BS.

Due to the existence of integer variables $\alpha_{ij} \in \{0,1\}$, $\beta_{jk} \in \{0,1\}$ and the continuous variables $ p_{ijk} \in [ 0, P_{\max} ] $, $\theta_m \in [0,2\pi]$, as well as their highly coupling in the non-convex objective function and constraints, It can be observed that the sum-rate maximization problem (7) is a mixed-integer non-linear programming (MINLP) problem, which is NP-hard \cite{Cui2018Optimal} and is non-trivial to solve optimally by common standard optimization methods.
Additionally, the exhaustive search method is not feasible, since the computational complexity grows exponentially over the total number of variables.
Therefore, it is essential to transform (7) into some tractable convex subproblems.

\section{Joint Optimization of Power, Reflection \\ and Decoding Order}
Given user association, subchannel assignment, and decoding order, our objective is to solve the problem of power allocation and reflection matrix design, which is given by
\begin{subequations}
\begin{eqnarray}
&\max \limits_{\mathbf{p},\mathbf{\Theta}} & \sum \nolimits_{i} \sum \nolimits_{j} \sum \nolimits_{k} R_{ijk}, \\
&s.t. & (7c),(7d),(7e),
\end{eqnarray}
\end{subequations}
where $\mathbf{p}=\left\lbrace p_{ijk} | \forall i,j,k \right\rbrace $ is the power allocation profile.
In inequality (7c), due to the inter-cell interference $I_{ijk}^{\text{inter}}$ and $I_{\tilde{i}jk}^{\text{inter}}$ for user $i$ and $\tilde{i}$, respectively, it is intractable to solve this non-linear and non-convex problem (8) by standard convex optimization approaches.
Since each BS aims for maximizing their sum rate, they are expected to allocate as much power as possible to their associated users.
Thus, we assume that the inter-cell interference $I_{ijk}^{\text{inter}}$ and $I_{\tilde{i}jk}^{\text{inter}}$ are approximately equal to the preset threshold $I_{\text{th}}$ for user $i$ and $\tilde{i}$ in the same cell, then $\Delta_{jk}(i,\tilde{i})$ is simplified as $|H_{\tilde{i}jk}|^2 \ge |H_{ijk}|^2$. 
Moreover, due to the non-concavity of $R_{ijk}$, we introduce an auxiliary variable set $\boldsymbol{\gamma} = \left\lbrace \gamma_{ijk} | \text{SINR}_{ijk} \ge \gamma_{ijk}, \forall i,j,k \right\rbrace $, and thus the problem (8) can be reformulated as
\begin{subequations}
\begin{eqnarray}
&\max \limits_{\mathbf{p},\mathbf{\Theta},\boldsymbol{\gamma}} & \sum \nolimits_{i} \sum \nolimits_{j} \sum \nolimits_{k} \frac{W}{K} \log_{2} \left( 1 + \gamma_{ijk} \right), \\
&s.t. & |H_{\tilde{i}jk}|^2 \ge |H_{ijk}|^2, \ \text{if} \ \pi_{jk}(i) \le \pi_{jk}(\tilde{i}), \\
&{}& \sum \nolimits_{j} \sum \nolimits_{k} \frac{W}{K} \log_{2} \left( 1 + \gamma_{ijk} \right) \ge R_{\text{min}}, \ \forall i, \\
&{}& \text{SINR}_{ijk} \ge \gamma_{ijk}, \ \forall i,j,k, \\
&{}& (7e).
\end{eqnarray}
\end{subequations}

\subsection{Power Allocation}
Given reflection matrix, the power allocation subproblem is given by
\begin{subequations}
\begin{eqnarray}
&\max \limits_{\mathbf{p},\boldsymbol{\gamma}} & \sum \nolimits_{i} \sum \nolimits_{j} \sum \nolimits_{k} \frac{W}{K} \log_{2} \left( 1 + \gamma_{ijk} \right), \\
&s.t. & (7e), (9c), (9d). 
\end{eqnarray}
\end{subequations}

Notice that all constraints in problem (10) are convex excluding constraint (9d), which can be recalculated as
\begin{equation}
\label{nonconvex_inequality}
p_{ijk} \ge
\gamma_{ijk} \hat{p}_{ijk} + \gamma_{ijk} \xi_{ijk},
\end{equation}
where $\hat{p}_{ijk} = \sum_{ \pi_{jk}(\hat{i}) > \pi_{jk}(i) } p_{\hat{i}jk}$
and $\xi_{ijk} = \frac{I_{\text{th}} + \sigma ^2}{|H_{ijk}|^2}$.
It is worth noting that the product term $\gamma_{ijk} \hat{p}_{ijk}$ is non-convex on the defined domain, $\gamma_{ijk} \ge 0, \hat{p}_{ijk} \ge 0$, and thus inequality (\ref{nonconvex_inequality}) is not a convex constraint.
Therefore, it is necessary to transform constraint (\ref{nonconvex_inequality}) into a convex one.

Let $f(\gamma_{ijk},\hat{p}_{ijk}) = \gamma_{ijk} \hat{p}_{ijk}$, while $\gamma_{ijk}, \hat{p}_{ijk} \ge 0$.
By replacing $f(\gamma_{ijk},\hat{p}_{ijk})$ with its convex upper bound (CUB), the resulting constraint becomes convex.
To this end, we define the following function:
\begin{equation}
\label{CUB}
g(\gamma_{ijk},\hat{p}_{ijk},\lambda_{ijk}) = \frac{\lambda_{ijk}}{2} \gamma_{ijk}^2 + \frac{1}{2\lambda_{ijk}} \hat{p}_{ijk}^2,
\end{equation}
where $\boldsymbol{\lambda} = \{\lambda_{ijk} | \forall i,j,k\}$ is a coefficient set. It can be proved that (\ref{CUB}) is a convex function, and $g(\gamma_{ijk},\hat{p}_{ijk},\lambda_{ijk}) \ge f(\gamma_{ijk},\hat{p}_{ijk})$ is satisfied for all $\lambda_{ijk}>0$.
Moreover, it can be derived that the equation will turn to equality when $\lambda_{ijk} = {\hat{p}_{ijk}}/{\gamma_{ijk}}$.
Hence, by replacing $f(\gamma_{ijk},\hat{p}_{ijk})$ with its convex upper bound $g(\gamma_{ijk},\hat{p}_{ijk},\lambda_{ijk})$, constraint (\ref{nonconvex_inequality}) is transformed as the following second-order cone constraint:
\begin{equation}
\label{SOC}
p_{ijk} \ge
\frac{\lambda_{ijk}}{2} \gamma_{ijk}^2 + \frac{1}{2\lambda_{ijk}} \hat{p}_{ijk}^2 + \gamma_{ijk} \xi_{ijk}.
\end{equation}

Next, by replacing (9d) with its approximate constraint (\ref{SOC}), it can be observed that both the objective function and all constraints in (10) becomes convex, and hence the Karush-Kuhn-Tucker (KKT) solution of (10) can be iteratively updated until convergence by optimally solving its convex approximation problem with CVX.
The details of the proposed CUB-based power allocation algorithm with adjustable convergence accuracy are summarized in Algorithm \ref{algorithm_1}, where the fixed coefficient $\gamma_{ijk}$ in the $n_{1}$-th iteration can be updated by
\begin{equation}\label{update_lambda}
\lambda_{ijk}^{(n_{1})} := {\hat{p}_{ijk}^{(n_{1}-1)}}/{\gamma_{ijk}^{(n_{1}-1)}}.
\end{equation}

\begin{algorithm}[h]
	\caption{CUB-Based Algorithm for Power Allocation}
	\label{algorithm_1}
	\begin{algorithmic}[1]
		\renewcommand{\algorithmicrequire}{\textbf{Initialize}}
		\renewcommand{\algorithmicensure}{\textbf{Output}}
		\STATE \textbf{Initialize} $\mathbf{p}^{(0)}$, $\boldsymbol{\gamma}^{(0)}$, the tolerance $\epsilon$, maximum iteration number $N_{1}$, and set current iteration number as $n_1=0$.
		\STATE Compute utility $U^{(0)} = \sum_{i} \sum_{j} \sum_{k} \frac{W}{K} \log_{2} ( 1 + \gamma_{ijk}^{(0)} )$;
		\REPEAT
		\STATE With given $\mathbf{p}^{(n_{1})}$ and $\boldsymbol{\gamma}^{(n_{1})}$, update $\boldsymbol{\lambda}^{(n_{1}+1)}$ by (\ref{update_lambda});
		\STATE With given $\boldsymbol{\lambda}^{(n_{1}+1)}$, obtain $\mathbf{p}^{(n_{1}+1)}$ and $\boldsymbol{\gamma}^{(n_{1}+1)}$ by solving the substituted problem of (10);
		\STATE With given $\boldsymbol{\gamma}^{(n_{1}+1)}$, compute $U^{(n_{1}+1)}=U( \boldsymbol{\gamma}^{(n_{1}+1)} ) $;
		\STATE Update $n_{1} := n_{1} + 1$;
		\UNTIL $|U^{(n_{1})}-U^{(n_{1}-1)}| < \epsilon $ or $n_{1} > N_{1}$;
		\STATE \textbf{Output} the converged solutions $\mathbf{p}^*$ and $\boldsymbol{\gamma}^{*}$;
	\end{algorithmic}
\end{algorithm}


\subsection{Reflection Matrix Design}
With the converged results $\mathbf{p}^*$ and $\boldsymbol{\gamma}^{*}$ derived from Algorithm 1, the problem (9) is simplified to the following feasibility-check subproblem:
\begin{subequations}
	\begin{eqnarray}
	&\text{find} & \mathbf{\Theta}, \\
	&s.t. & (9b),(9d).
	\end{eqnarray}
\end{subequations}

For notational convenience, we define $\boldsymbol{\rho}_{ijk} = \text{diag}\{ \mathbf{g}_{ik}^{\rm H} \} \mathbf{f}_{jk}$ and $\boldsymbol{\nu} = [\nu_1, \nu_2, \dots, \nu_M]^{\rm H}$, where $\nu_m = e^{j\theta_m}$. Thus, $|h_{ijk} + \mathbf{g}_{ik}^{\rm H} \mathbf{\Theta} \mathbf{f}_{jk}| = |h_{ijk} + \boldsymbol{\nu}^{\rm H} \boldsymbol{\rho}_{ijk} |$.
Meanwhile, we denote the real and imaginary parts of $H_{ijk}$ as $x_{ijk}$ and $y_{ijk}$, respectively, such that $x_{ijk}^2 + y_{ijk}^2 = |h_{ijk} + \boldsymbol{\nu}^{\rm H} \boldsymbol{\rho}_{ijk}|^2$.
Then, the feasibility-check problem (15) is rewritten as
\begin{subequations}
	\begin{IEEEeqnarray}{ll}	
		&\text{find} \quad \boldsymbol{\nu}, \\
		s.t.\ & x_{\tilde{i}jk}^2 + y_{\tilde{i}jk}^2 \ge x_{ijk}^2 + y_{ijk}^2, \text{if} \ \pi_{jk}(i) \le \pi_{jk}(\tilde{i}), \\
		{}& x_{ijk}^2 + y_{ijk}^2 \ge (x_{ijk}^2 + y_{ijk}^2) \phi_{ijk} + \hat{\xi}_{ijk}, \ \\
		{}& | \nu_m | = 1, \ \forall m, \\
		{}& x_{ijk} = \text{real} \left( h_{ijk} + \boldsymbol{\nu}^{\rm H} \boldsymbol{\rho}_{ijk} \right), \\
		{}& y_{ijk} = \text{imag} \left( h_{ijk} + \boldsymbol{\nu}^{\rm H} \boldsymbol{\rho}_{ijk} \right),
	\end{IEEEeqnarray}
\end{subequations}
where $\phi_{ijk} = \frac{\gamma_{ijk} \hat{p}_{ijk}}{p_{ijk}}$ and $\hat{\xi}_{ijk} = \frac{\gamma_{ijk} (I_{\text{th}} + \sigma^2)}{p_{ijk}}$.
Owing to the non-convex constraints (16b) and (16c), the problem (16) is non-trivial to be solved directly.
Thus, we invoke successive convex approximation (SCA) to replace $x_{ijk}^2 + y_{ijk}^2$ with its first-order Taylor approximation and iteratively solve the resulting problem until it converges to a KKT solution within the preset accuracy.
Toward this end, the lower-bound approximation for $x_{ijk}^2 + y_{ijk}^2$ is given by
\begin{eqnarray}
\tau(x_{ijk},y_{ijk}) &=& \tilde{x}_{ijk}^2 + \tilde{y}_{ijk}^2 + 2 \tilde{x}_{ijk} (x_{ijk}-\tilde{x}_{ijk}) \nonumber \\ 
&+& 2 \tilde{y}_{ijk} (y_{ijk}-\tilde{y}_{ijk}),
\end{eqnarray}
where $\{(\tilde{x}_{ijk},\tilde{y}_{ijk}) | \forall i,j,k\}$ is a set of feasible solution of (16), and they can be updated in the $n_2$-th iteration by
\begin{eqnarray}
\label{updat_x}
&{}& \tilde{x}_{ijk}^{(n_2)} := \text{real} \left( h_{ijk} + \boldsymbol{\nu}^{\rm H} \boldsymbol{\rho}_{ijk}^{(n_2 - 1)} \right), \\
\label{updat_y}
&{}& \tilde{y}_{ijk}^{(n_2)} := \text{imag} \left( h_{ijk} + \boldsymbol{\nu}^{\rm H} \boldsymbol{\rho}_{ijk}^{(n_2 - 1)} \right).
\end{eqnarray}

Afterwards, by replacing the expressions on the left-side of (16b) and (16c) with their first-order Taylor approximation, it can be observed that the substituted problem of (16) becomes a convex one, which also can be solved with CVX.


\subsection{Decoding Order Determination}
The decoding order in each cell depends on the combined channel gains experienced by users clustered in the cell.
Due to the same phase shifts applied for all users with different channels, the combined channel gains of different users cannot be maximized at the same time.
Thus, we alternatively maximize the sum of all combined channel gains, which is recalculated as
\begin{subequations}
	\begin{eqnarray}
	&\max \limits_{\mathbf{\Theta}}& \sum \nolimits_{i} \sum \nolimits_{j} \sum \nolimits_{k} |H_{ijk}|^2, \\
	&s.t.& \theta_m \in [0,2\pi], \ \forall m.
	\end{eqnarray}
\end{subequations}

Define $ \boldsymbol{\varUpsilon}_{ijk} = \text{real}(\boldsymbol{\nu}^{\rm H} \boldsymbol{\rho}_{ijk})$ and $\boldsymbol{\Gamma}_{ijk} = \boldsymbol{\rho}_{ijk} \boldsymbol{\rho}_{ijk}^{\rm H}$, then it can be noticed that
\begin{equation}
|h_{ijk} + \boldsymbol{\nu}^{\rm H} \boldsymbol{\rho}_{ijk}|^2 = |h_{ijk}|^2 + 2 h_{ijk} \boldsymbol{\varUpsilon}_{ijk} + \boldsymbol{\nu}^{\rm H} \boldsymbol{\Gamma}_{ijk} \boldsymbol{\nu}.
\end{equation}

Therefore, we have
\begin{equation}
|H_{ijk}|^2 = \boldsymbol{\bar{\nu}}^{\rm H} \mathbf{C}_{ijk} \boldsymbol{\bar{\nu}} + |h_{ijk}|^2,
\end{equation}
where
\begin{equation}
\mathbf{C}_{ijk} = \left[ \begin{array}{cc}
\boldsymbol{\Gamma}_{ijk} & h_{ijk} \boldsymbol{\rho}_{ijk} \\
h_{ijk} \boldsymbol{\rho}_{ijk}^{\rm H} & 0 
\end{array} 
\right ]
\text{and} \
\boldsymbol{\bar{\nu}} = \left[ \begin{array}{c}
\boldsymbol{{\nu}} \\
1
\end{array} 
\right ].
\end{equation}

Furthermore, we define $\textbf{V} = \boldsymbol{\bar{\nu}} \boldsymbol{\bar{\nu}}^{\rm H}$, while $\textbf{V} \succeq \mathbf{0}$ and $\text{rank}(\textbf{V}) = 1$.
Then we have $\boldsymbol{\bar{\nu}}^{\rm H} \mathbf{C}_{ijk} \boldsymbol{\bar{\nu}} = \text{tr}(\mathbf{C}_{ijk} \textbf{V})$, and the problem (20) is equivalently reformulated as 
\begin{subequations}
	\begin{eqnarray}
	&\max \limits_{\mathbf{V}} & \sum \nolimits_{i} \sum \nolimits_{j} \sum \nolimits_{k} \text{tr}(\mathbf{C}_{ijk} \textbf{V}) + |h_{ijk}|^2, \\
	&s.t.& V_{m,m} = 1, \ m = 1,2, \ldots, M+1 \\
	&{}& \textbf{V} \succeq \mathbf{0}, \\
	&{}& \text{rank}(\textbf{V}) = 1.
	\end{eqnarray}
\end{subequations}

Although the rank-one constraint is still non-convex, the semidefinite relaxation (SDR) can be applied to relax (24) to a convex SDP problem, and thus the optimal $\mathbf{V}^{*}$ can be obtained by solving the relaxed problem with CVX.
Finally, with $\mathbf{V}^{*} = \boldsymbol{\bar{\nu}}^{*} \boldsymbol{\bar{\nu}}^{*H}$, the optimal reflection matrix $\mathbf{\Theta}^{*}$ is obtained.
Based on the converged results $\mathbf{p}^*$ and $\mathbf{\Theta}^{*}$, if the combined channel gains experienced by any two users $(i, \tilde{i})$ in each cell $j$ on subchannel $k$ can be arranged as $H_{ijk} \le H_{\tilde{i}jk}$, then the decoding order is given by $\pi_{jk}(i) \le \pi_{jk}(\tilde{i})$.
However, if $\text{rank}(\textbf{V}) \ne 1$, then objective value obtained from the relaxed problem is only an upper bound of (20).
Thus, the Gaussian randomization (GR) method can be invoked to construct a rank-one solution based on the higher-rank solution of the relaxed problem, which is omitted for space reason.

\section{Matching Theory for User Association \\ and Subchannel Assignment}
In this section, we focus on the user association and subchannel assignment problem in (7) with fixed power allocation strategy and reflection matrix, which can be expressed as
\begin{subequations}
	\label{3D_matching}
	\begin{eqnarray}
	\label{matching_problem}
	&\max \limits_{\boldsymbol{\alpha},\boldsymbol{\beta}} & \sum_{i=1}^{I} \sum_{j=1}^{J} \sum_{k=1}^{K} R_{ijk}, \\
	&s.t.& (7f)-(7i),
	\end{eqnarray}
\end{subequations}
where $\boldsymbol{\alpha}=\{\alpha_{ij} | \forall i,j\}$ denotes the user association profile and $\boldsymbol{\beta} = \{ \beta_{jk} | \forall j,k\}$ represents the subchannel assignment profile.
It can be observed that (\ref{3D_matching}) is a 3D matching problem.
In order to address this challenging issue, we decompose the 3D matching problem into two 2D matching problems, i.e., user association problem and subchannel assignment problem.
%
Note that the decomposed 2D matching problem is a many-to-many(one) matching problem with peer effects.

During the matching process, each player $e \in \mathcal{E}$ has a transitive and strict preference list with respect to its interests over the set of $\mathcal{W}$, and vice versa.
We use $w_1 \succ_{e} w_2$ to denote that player $e$ strictly prefers $w_1$ to $w_2$.
If $w_2 \succ_{e} w_3$ is satisfied at the same time, then we have $w_1 \succ_{e} w_3$.
Given a matching function $\mu$, and assume that $\mu(e)=w$ and $\mu(e')=w'$.
Then,
in order to handle the peer effects and ensure exchange stability, 
we define the swap matching as
\begin{equation}
\label{swap_matching}
\mu_{e}^{e'} = \left\lbrace \mu \backslash \{ (e,w),(e',w')\} \bigcup \{(e',w),(e,w')\} \right\rbrace,
\end{equation}
where players $e$ and $e'$ exchange their matched players $w$ and $w'$ while keeping all other matching state the same.
Based on the swap operation in (\ref{swap_matching}), we define the concept of swap-blocking pair as follows.

\begin{definition}
	A pair of players $(e,e')$ is called a swap-blocking pair in $\mu$ if and only if 
1) $\forall q \in \{e,e',w,w'\}$, $U_q(\mu_{e}^{e'}) \ge U_q(\mu)$, and
2) $\exists q \in \{e,e',w,w'\}$, such that $U_q(\mu_{e}^{e'}) > U_q(\mu)$,
	where $U_q(\mu)$ denotes the utility value (i.e., achievable data rate) of player $q$ under the matching $\mu$.
	It is worth noting that the matching $\mu$ is two-sided exchange-stable if and only if there dose not exist a swap-blocking pair.
\end{definition}


\subsection{Many-to-One Matching for User Association}
\label{5-A}
In the many-to-one matching problem of user association, we define the preference of user $i$ associated with BS $j$ as
$
U_{ij} = \sum \nolimits_{k \in \mathcal{K}} \frac{W}{K} \log_{2} \left( 1+\gamma_{ijk} \right).
$
If user $i$ can achieve a higher data rate when being associated with BS $j$ compared to be that of being associated with BS $j'$, i.e., user $i$ prefers to be associated with BS $j$ in matching $\mu$ rather than BS $j'$ in matching $\mu'$, then we have 
\begin{equation}
\label{user_preference}
(j,\mu) \succ_{i} (j',\mu') \ \Leftrightarrow \ U_{ij}(\mu) > U_{ij'}(\mu').
\end{equation}

Similarly, the preference of BS $j$ associated with user set $\mu(j)$ is given by
$
U_{j} = \sum \nolimits_{i \in \mu(j)} \sum \nolimits_{k \in \mathcal{K}} \frac{W}{K} \log_{2} \left( 1+\gamma_{ijk} \right).
$
For any two subsets of users $\mathcal{I}_1=\mu(j)$ and $\mathcal{I}_2=\mu'(j)$ while $\mathcal{I}_1 \ne \mathcal{I}_2$, if BS $j$ obtain get a higher data rate when being associated with $\mathcal{I}_1$ than that of being associated to $\mathcal{I}_2$, i.e., BS $j$ prefers user subset $\mathcal{I}_1$ in matching $\mu$ to user subset $\mathcal{I}_2$ in matching $\mu'$, then we have
\begin{equation}
\label{BS_preference}
(\mathcal{I}_1,\mu) \succ_{j} (\mathcal{I}_2,\mu') \ \Leftrightarrow \ U_{j}(\mu) > U_{j}(\mu').
\end{equation}

According to (\ref{user_preference}) and (\ref{BS_preference}), the preference lists of all users and BSs are constructed.
Subsequently, each user proposes to the most preferred BS that has never rejected them before.
Then, each BS accepts the most preferred users and rejects the others.
Finally, the initial matching state between users and BSs is obtained when there is no unmatched user.
After that, each user tries to search for another user to form a swap-blocking pair and swaps their matching states based on (\ref{swap_matching}), which terminates when no swap-blocking pair exists.
In summary, the many-to-one matching for user association is described in Algorithm \ref{algorithm_4}.

\begin{algorithm}
	\caption{Many-to-One Matching for User Association}
	\label{algorithm_4}
	\begin{algorithmic}[1]
		\renewcommand{\algorithmicrequire}{\textbf{Initialize}}
		\renewcommand{\algorithmicensure}{\textbf{Output}}
		\STATE \textbf{Initialize} the User-BS matching state as $\Phi_{1}$.
		\REPEAT
		\STATE For every user $i \in \Phi_{1}$, it searches for another user $i' \in \Phi_{1} \backslash \Phi_{1}(\mu(i))$ to check whether $(i,i')$ is a swap-blocking pair;
		\IF {$(i,i')$ is a swap-blocking pair}
		\STATE Update $\mu := \mu_{i}^{i'}$;
		\ELSE
		\STATE Keep the current matching state;
		\ENDIF
		\UNTIL No swap-blocking pair can be constructed.
		\STATE \textbf{Output} the stable User-BS matching $\mu^{*}$ and its corresponding utility $U_1=U(\mu^{*})$.
	\end{algorithmic}
\end{algorithm}

\subsection{Many-to-Many Matching for Subchannel Assignment}
Analogously, in the many-to-many matching problem of subchannel assignment, the preference of BS $j$ on subchannel $k$ is defined as
$
U_{jk} = \sum \nolimits_{i \in \mathcal{I}} \frac{W}{K} \log_{2} \left( 1+\gamma_{ijk} \right).
$
If BS $j$ can achieve a higher data rate when being assigned with subchannel $k$ compared to that of being assigned with subchannel $k'$, i.e., BS $j$ prefers to subchannel $k$ in matching $\mu$ rather than subchannel $k'$ in matching $\mu'$, then we have 
\begin{equation}
\label{user_BS_preference}
(k,\mu) \succ_{j} (k',\mu') \ \Leftrightarrow \ U_{jk}(\mu) > U_{jk'}(\mu').
\end{equation}

Similarly, the preference of subchannel $k$ on BS set $\mu(k)$ is given by
$
U_{k} = \sum \nolimits_{i \in \mathcal{I}} \sum \nolimits_{j \in \mu(k)} \frac{W}{K} \log_{2} \left( 1+\gamma_{ijk} \right).
$
For any two subsets of BSs $\mathcal{J}_1=\mu(k)$ and $\mathcal{J}_2=\mu'(k)$ while $\mathcal{J}_1 \ne \mathcal{J}_2$, if subchannel $k$ can get a higher data rate when being assigned to $\mathcal{J}_1$ than $\mathcal{J}_2$, i.e., subchannel $k$ prefers to BS subset $\mathcal{J}_1$ in matching $\mu$ rather than BS subset $\mathcal{J}_2$ in matching $\mu'$, then we have
\begin{equation}
\label{subchannel_preference}
(\mathcal{J}_1,\mu) \succ_{k} (\mathcal{J}_2,\mu') \ \Leftrightarrow \ U_{k}(\mu) > U_{k}(\mu').
\end{equation}

First, the preference lists of all (User-BS) units and subchannels are established according to (\ref{user_BS_preference}) and (\ref{subchannel_preference}).
Then, an initial matching state can be generated by adopting the aforementioned method in Section \ref{5-A}.
Finally, the search process is executed based on (\ref{swap_matching}), which terminates until there exists no swap-blocking pair.
The many-to-many matching for subchannel assignment is described in Algorithm \ref{algorithm_5}.

\begin{algorithm}
	\caption{\small Many-to-Many Matching for Subchannel Assignment}
	\label{algorithm_5}
	\begin{algorithmic}[1]
		\renewcommand{\algorithmicrequire}{\textbf{Initialize}}
		\renewcommand{\algorithmicensure}{\textbf{Output}}
		\STATE \textbf{Initialize} (User,BS)-Subchannel matching state as $\Phi_{2}$. 
		\REPEAT
		\STATE For every (User,BS) $j \in \Phi_{2}$, it searches for another (User,BS) $j' \in \Phi_{2} \backslash \Phi_{2}(\mu(j))$, and let $\mathcal{U}=\{U_1\}$;
		\STATE For a given $j$, calculate the candidate $U_{j}^{j'}$ for the swapping pair $(j,j')$;
		\IF {$(j,j')$ is a swap-blocking pair}
		\STATE Update $\mathcal{U} := \mathcal{U} \cup \{U_{j}^{j'}\}$;
		\ENDIF
		\STATE Find $j^{'*} = \arg \max_{j'} \mathcal{U}$;
		\STATE Update $\bar{\mu} := \bar{\mu}_{j}^{j^{'*}}$, and set $U_2 = U_{j}^{j^{'*}}$;
		\UNTIL No swap-blocking pair can be constructed.
		\STATE \textbf{Output} the stable (User,BS)-Subchannel matching $\bar{\mu}^{*}$.
	\end{algorithmic}
\end{algorithm}

\begin{figure*}[t]
	\centering
	\begin{minipage}[t]{0.32 \textwidth}
		\centering
		\includegraphics[width=2.3 in]{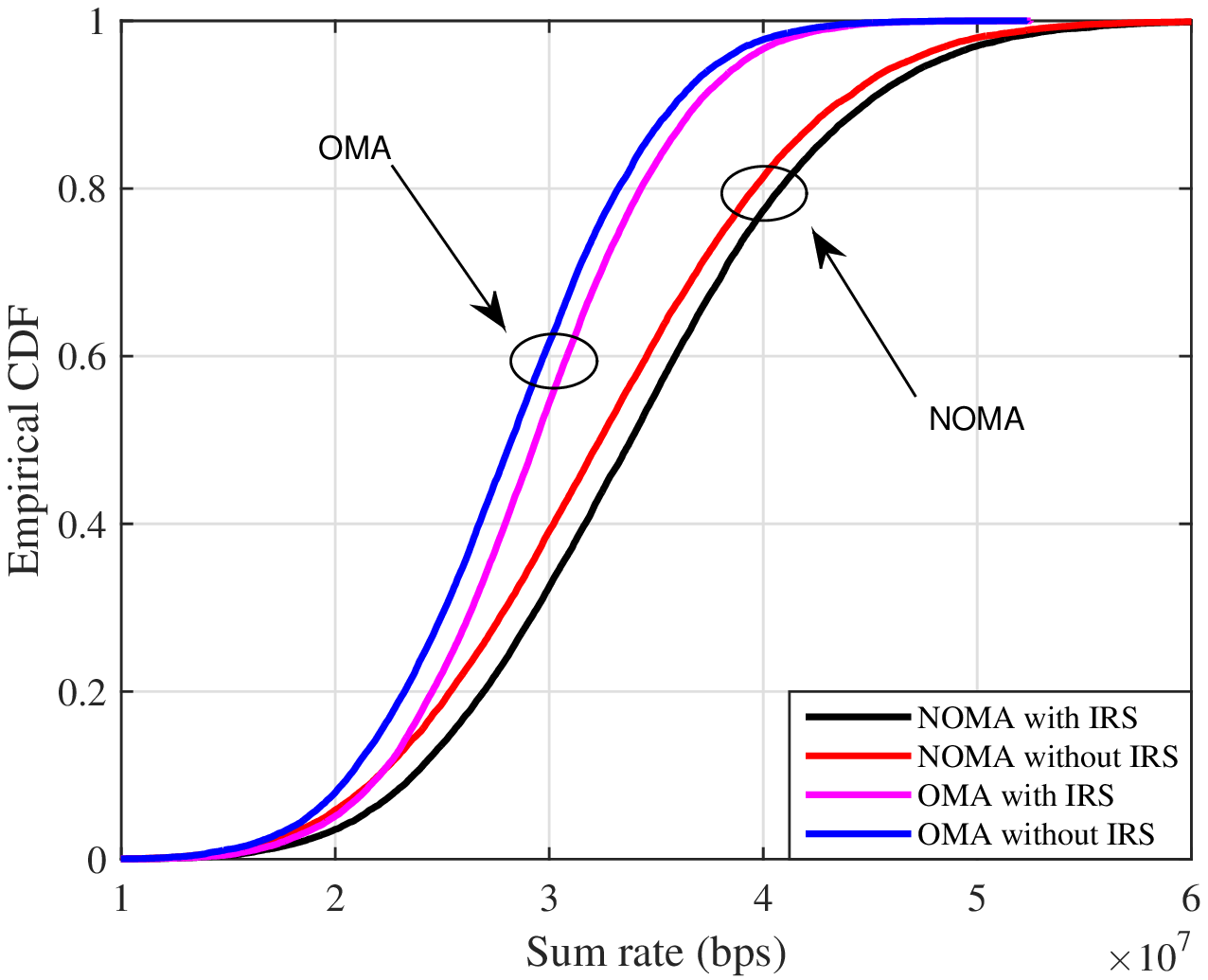}
		\caption{Empirical CDF of sum rate.}
		\label{fig2}
	\end{minipage}
	\begin{minipage}[t]{0.32 \textwidth}
		\centering
		\includegraphics[width=2.3 in]{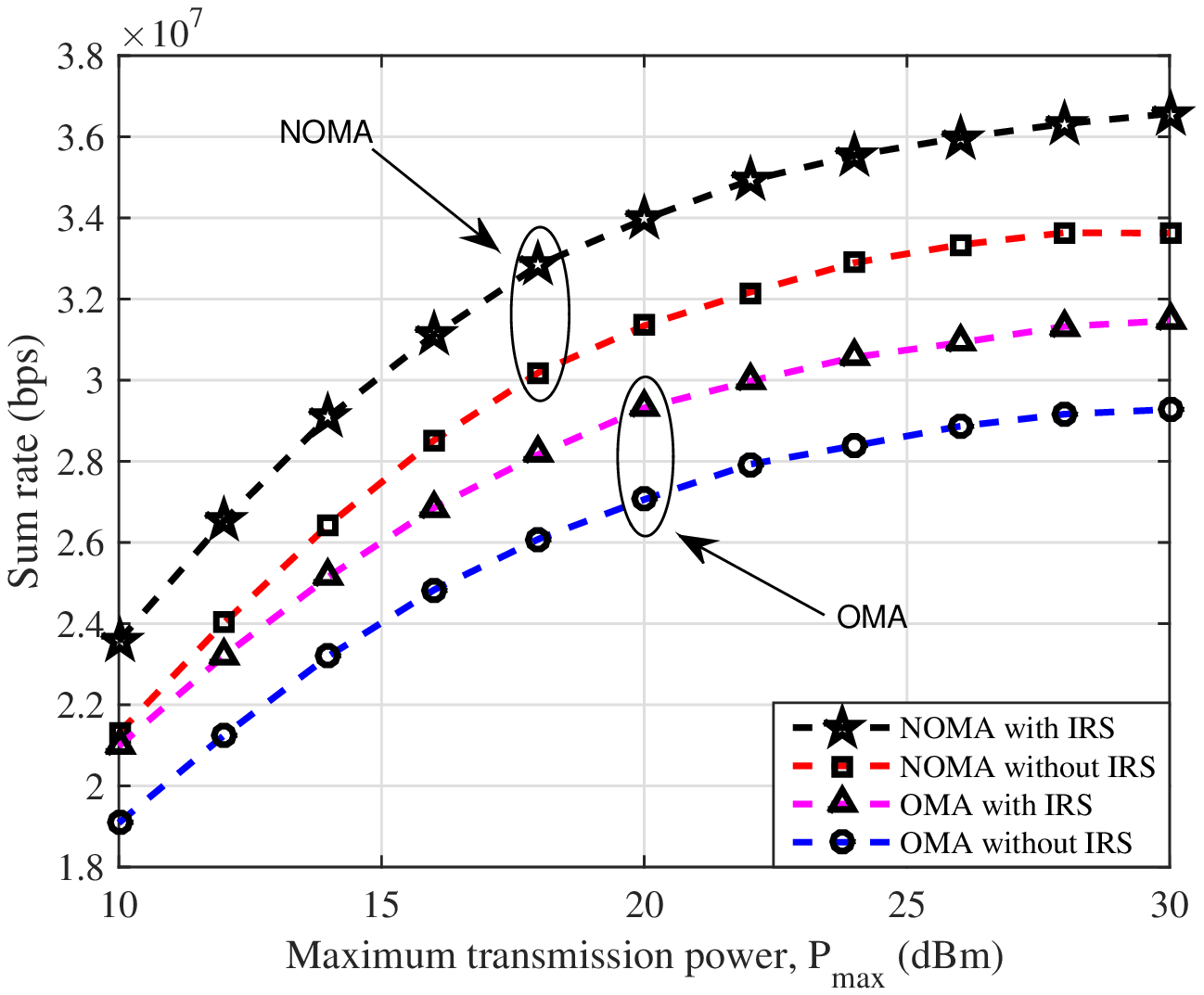}
		\caption{Sum rates versus $P_{\text{max}}$.}
		\label{fig3}
	\end{minipage}
	\begin{minipage}[t]{0.32 \textwidth}
		\centering
		\includegraphics[width=2.3 in]{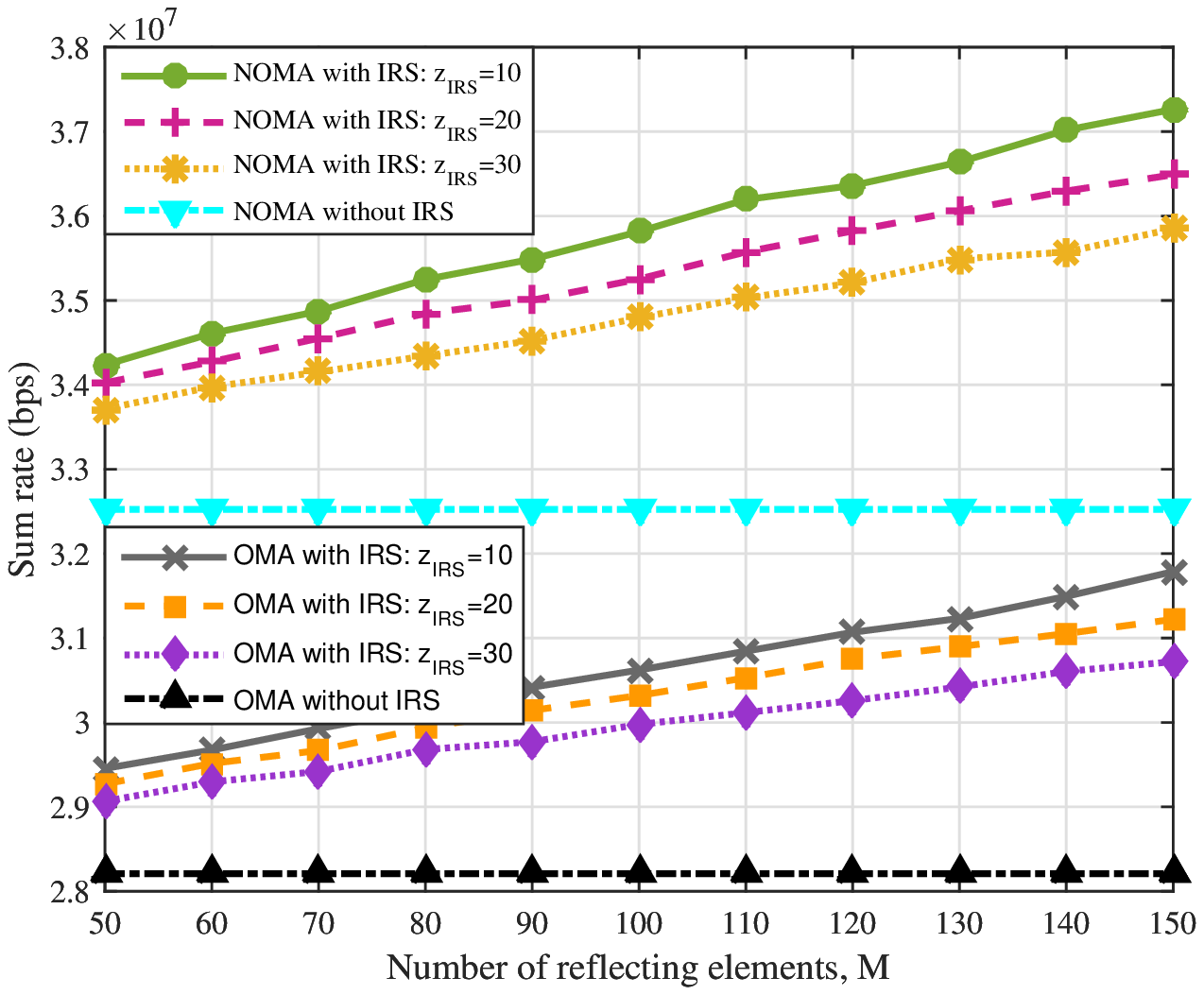}
		\caption{Sum rates versus M.}
		\label{fig4}
	\end{minipage}
\end{figure*}

\section{Numerical Simulation}
We consider that there are $6$ users, $3$ BSs and $3$ subchannels in the multi-cell IRS-aided NOMA network, where user $i$, BS $j$ and IRS are located at $(x_i,y_i,z_i) = (50i,30,0)$, $(x_j,y_j,z_j) = (100j,0,20)$ and $(200,50,20)$, respectively.
Moreover, the number of reflecting elements is set as $M=100$, the system bandwidth is assumed to be $W=3$ MHz.
The noise power is $\sigma^2 = -80$ dBm, and the minimum rate requirement of each user is assumed to be $R_{\text{min}}=500$ Kbps.
The maximum transmission power at each BS is set as $P_{\text{max}} = 23$ dBm, unless otherwise stated.
We simulate 2000 runs, all results are averaged over independent realization.


%
%

%

In Fig. \ref{fig2}, it can be observed that the NOMA schemes enjoy a significant performance gain than OMA schemes, which is mainly because NOMA allows multiple users to reuse the same subchannel, and thus can obtain a higher spectrum efficiency.
In particular, the IRS-aided NOMA/OMA networks can achieve better performance than the conventional NOMA/OMA schemes without IRS in terms of achievable sum rate, which demonstrates that IRS is capable of enhancing the system performance by proactively modifying the wireless channel between the transmitter and receiver.

In Fig. \ref{fig3}, it can be seen that the achievable sum rate of all the four schemes increase when the maximum transmission power increases.
One can notice that the lower the $P_{\text{max}}$ value is, the larger the slope of the sum rate curves will be.
Thus, different from the approximately linear growth at a low $P_{\text{max}}$, the sum rate curves increase more slowly at a high $P_{\text{max}}$ due to the existence of intra-cell and inter-cell interference.
It’s worth pointing that the performance of NOMA/OMA schemes would reach their peak as the maximum transmission power increases to a certain threshold.

In Fig. \ref{fig4}, it can be observed that the sum rate achieved by the IRS-aided NOMA/OMA networks approximately linear increase over the number of reflecting elements, which significantly outperform that of the benchmark schemes without IRS.
This indicates that the wireless environment is more controllable and programmable in the IRS-aided networks with more reflecting elements.
Thus, better performance can be achieved by employing a large number of reflecting elements to alleviate interferences and enhance the desired signals.

\section{Conclusion}
In this paper, we investigated the sum-rate maximization problem in the IRS-aided multi-cell NOMA network, which was formulated as a MINLP problem.
Then, relaxation methods were invoked to transform the intractable subproblems into convex ones, and efficient algorithms were designed to solve these challenging subproblems iteratively.
Next, in order to achieve a two-sided exchange-stable state among users, BSs and subchannels, swap matching-based algorithms were proposed.
Finally, numerical results showed that through reconfiguring the wireless environment, IRS is capable of enhancing system performance,
and the proposed algorithms can improve both the throughput and energy efficiency.

%
%
%
%
%
%
%

\bibliographystyle{IEEEtran}
\bibliography{IEEEabrv,ref}

\end{document}